\documentclass[preprint2]{aastex}

\shorttitle{Study of errors in strong gravitational lensing}
\shortauthors{Frittelli and Kling}

\begin{document}

\title{Study of errors in strong gravitational lensing}

\author{Thomas P. Kling}  \email{tkling@bridgew.edu}

\affil{Department of Physics, Bridgewater State College,
Bridgewater, MA 02325}

\author{Simonetta Frittelli}

\affil{Department of Physics, Duquesne University, Pittsburgh, PA
15282}


\begin{abstract} We examine the accuracy of strong gravitational
lensing determinations of the mass of galaxy clusters by comparing
the conventional approach with the numerical integration of the fully
relativistic null geodesic equations in the case of weak
gravitational perturbations on Robertson-Walker metrics.  In
particular, we study spherically-symmetric, three-dimensional
singular isothermal sphere models and the three-dimensional matter
distribution of \citet{NFW}, which are both commonly used in
gravitational lensing studies.  In both cases we study two different
methods for mass-density truncation along the line of sight: hard
truncation and conventional (no truncation). We find that the
relative error introduced in the total mass by the thin lens
approximation alone is less than $0.3\%$ in the singular isothermal
sphere model, and less than $2\%$ in the model of \citet{NFW}.  The
removal of hard truncation introduces an additional error of the same
order of magnitude in the best case, and up to an order of magnitude
larger in the worst case studied. Our results ensure that the future
generation of precision cosmology experiments based on lensing
studies will not require the removal of the thin-lens assumption,
but they may require a careful handling of truncation.

\end{abstract}




\section{Introduction} \label{intro:sec}

In the current age of precision cosmology, the fundamental parameters
of the favored cosmological model can be measured with high
accuracy.  Strong and weak gravitational lensing studies are
particularly important in this endeavor as an independent and
relatively assumption-free measure of mass morphology in clusters of
galaxies.

Conventional gravitational lensing studies continue to utilize the
standard assumption of thin lenses.  This allows for the actual path
of photons from source to observer to be approximated by its
asymptotics, with a sharp bending at the plane where the lens is
located.  There is, clearly, some error involved in this
approximation, but the size of this error has not been rigorously
examined, to our knowledge.

A question that will be important as the accuracy increases is at
what point the error in the conventional approach becomes comparable
to the uncertainty in the cosmological parameters.  At that point,
accurate modeling of lensing events  will require a more accurate
version of the lens equation, or an efficient algorithm to calculate
the photon paths. In this respect, seminal proposals that improve on
the conventional approach exist, dating from the 1990's.  In the
work of \citet{pyne}, higher-order lens equations in cosmological
backgrounds are developed.  A different style of approximation which
correctly utilizes the thin-lens paths as the zeroth-order path is
introduced in \citet{knp1,knp2}.

The same motivation sparked some interest in revisiting the
foundations of gravitational lensing theory with the aim of
developing the concept of the lens equation without reference to a
lens plane, as in \citet{efn,fn} and \citet{perlick}.  These works
develop a consistent theory of strong lensing entirely on the basis
of the null geodesics in an arbitrary spacetime. Image distortion and
weak gravitational lensing were considered from this perspective in
\citet{fkn2,fkn3}. An application of this no-lens-plane approach in
(non-cosmological) spacetimes representing spherically symmetric
non-singular matter distributions is presented in \citet{kn}. The
case of Schwarzschild black holes is developed in \citet{fkn1} and
the resulting lensing predictions are compared to \citet{ve}, where a
lens-plane approximation is applied to lightrays that undergo large
bending.  Large bending angles by static and spinning black holes are
also treated with an approximation scheme in  \citet{bz,bz1}, with an
application to Sag A$^*$.

The purpose of this paper is to test the accuracy of the
conventional approach in determining the total mass of clusters. For
this purpose we integrate the null geodesic equations in
cosmological lensing spacetimes. We work with actual or realistic
strong-lensing scenarios in the standard cosmology, assuming matter
distributions of either a three-dimensional singular isothermal
sphere (SIS) model or the model of \citet{NFW} (NFW). Both these
models are unrealistic to the extent that the mass density does not
fall off fast enough to allow for a bounded mass in all space. In
principle, some way of truncating the mass density is implicit in
the model, although the details of the truncation mechanism are not
considered of fundamental relevance to the predicted observables.

Often in the application of gravitational lensing to astrophysical
data, truncation is imposed only in the directions transverse to the
line of sight. This practice is justified on the basis that the
projected mass density required for the thin-lens approximation is
actually well defined in the model, as it involves only a one
dimensional integration of the mass density along the line of sight,
and therefore, truncation of the model along the line of sight is
not necessary. Naturally, this practice introduces further error in
addition to the error involved in the use of the thin-lens
assumption.  The accuracy of this conventional model (thin-lens plus
lack of truncation along the line of sight) is the issue in our
current study.

Our method consists of comparing the observables predicted by the
conventional approach to those predicted by the theory with no
approximations. The theory is provided by the bending of light in
the presence of mass as described by general relativity. The
physical scenario common to both problems is the mass-density model.
This is used differently in both approaches.  In the conventional
approach, the mass density is only used after integration along the
line of sight, with no truncation.  In the relativistic approach,
which here we consider to be the exact treatment of the problem, the
mass density is used as a source of the Newtonian potential that
determines the components of the metric whose null geodesics are
sought in exact form. The general relativistic (or exact) approach
does not make sense in the absence of truncation along the line of
sight.  Therefore, our study ends up comparing the results of a
non-truncated thin-lens model to a properly truncated
general-relativistic numerical integration, both of which represent
the same physical situation.  For a proper interpretation of our
results, it is important to emphasize that the non-truncated
thin-lens models are conventionally intended to represent, with some
accuracy, the physical situation of a properly truncated mass
density, and for this reason the comparison is entirely justified.

Clearly, however, the comparison of the conventional approach with
the relativistic treatment provides a measure of the total error
incurred by two separate mechanisms: the thin lens approximation,
and the lack of proper truncation along the line of sight.  The
error introduced by the truncation mechanism, or the lack of it,
into the thin-lens scheme can be analyzed as an end in itself. In
this respect, progress has been made recently in introducing better
physical truncation schemes that improve over hard truncation,
including \citet{baltz} and \citet{takada}. Of importance,
\citet{baltz} considers differences between truncated and
non-truncated lensing observables, giving a measure of the
inaccuracy introduced by the truncation mechanism into the thin-lens
scheme.  These studies take the properly truncated thin-lens scheme
as the exact model of the physical situation, thus they fail to
properly account for the actual inaccuracy with respect to the
general relativistic theory. The size of the truncation error by
itself within the thin-lens scheme has little meaning if the actual
error introduced by the thin-lens assumption is not known.

In order to give an idea of the sizes of both errors (thin-lens and
lack of truncation) with respect to the relativistic theory, we also
compare our integration of the null geodesics with thin-lens models
truncated along the line of sight.  In this case, the inaccuracy of
the model is entirely due to a single assumption: that of thin
lenses. Our mechanism of truncation is as simple as possible: a hard
cut off of the mass density after an arbitrarily chosen radius.

In Section~\ref{back:sec}, we outline the ways the Robertson-Walker
(RW) background geometry enters our calculation, and in
Section~\ref{eqns:sec}, we outline the equations of motion.  The SIS
and NFW gravitational potentials and thin-lens models are developed
in section~\ref{models:sec}.  Section~\ref{sim:sec} describes our
experimental set-up, numerical approach, and numerical accuracy. Our
main results comparing the accuracy of the thin-lens models are
presented in section~\ref{compare:sec}.  Section~\ref{cosmo:sec}
presents the variation of thin-lens accuracy as a function of
cosmology parameters for a range of flat cosmologies.  We discuss
and explain our results in section~\ref{discussion:sec} before
concluding with some comments about our results and the their
implications for lensing studies in section~\ref{conclusion:sec}.


\section{Cosmological background} \label{back:sec}

We take as the background a flat RW metric with the current accepted
values of the Hubble constant $H_0 = 70$~km/s/Mpc, the matter density
$\Omega_m = 0.3$ and the cosmological constant density $\Omega_\Lambda
= 0.7 = 1- \Omega_m$. The metric is

\begin{equation} ds^2 = c^2dt^2 - a^2(t) \left\{ dr^2 + r^2 (d\theta^2
+ \sin^2\theta d\phi^2) \right\}, \label{FRW1} \end{equation}

\noindent with a scale factor $a(t)$ for our cosmological model
as in \cite{ryden}

\begin{equation} a(t) = \left( \frac{\Omega_m}{\Omega_\Lambda}
\right)^{1/3} \left\{ \sinh\left( \frac{3 H_0 \sqrt{\Omega_\Lambda}
t}{2} \right) \right\}^{2/3}. \label{a1}  \end{equation}

\noindent The Hubble parameter $H\equiv \dot{a}/a$ is

\begin{equation} H^2 = H_0^2 \left( \frac{\Omega_m}{a(t)^3} + (1 -
\Omega_m) \right). \label{hubble} \end{equation}

We will be considering photons emitted by a source at emission time
$t_e$, and arriving at the observer at the observation time $t_o$.
For numerical convenience, we fix the scale factor to the value 1 at
the observation time, $a(t_o) = 1$, which yields $t_o = 4.248 \times
10^{17}\,s $.

A natural way to give the relative locations between the lens,
observer, and source is to place the lens at the spatial origin, and
to use the redshifts to determine the positioning of the lens and
observer.  Using the redshift relation, $ 1+\tilde z =
\frac{a(t_o)}{a(t_e)}, $ and setting $a(t_o)=1$, we can solve for
the value of $t_e$ with Eq.~\ref{a1}.

To obtain the radial positions of the source and observer, we orient
our coordinates in such a way that a light ray travels radially in
the background spacetime and assume that the observer, lens and
source are at least nearly co-linear. We can then integrate radial
null geodesics of the metric Eq.~\ref{FRW1} to determine the radial
positions of the source and observer (by ignoring the perturbation
introduced by the lens).


\section{Equations of motion} \label{eqns:sec}

The lensing scenario is described in terms of the following weakly
perturbed RW metric

\begin{eqnarray} ds^2 &=& (1+2 \varphi)dt^2
- a^2
(1-2\varphi) \nonumber\\
&&\times \{ dr^2 + r^2 (d\theta^2 + \sin^2\theta d\phi^2 )\} ,
\label{m1} \end{eqnarray}

\noindent This metric is accurate to first order in $\varphi$, which is a
Newtonian potential determined by the proper mass density of the lens
$\rho_p$ via

\begin{equation} \nabla_p^2 \varphi = 4 \pi G \rho_p, \label{poisson}
\end{equation}

\noindent where the derivatives in the Laplacian operator are taken
with respect to the proper distance $r_p = a(t) r$. Commonly, the
proper mass density is taken as a function $\rho_p(r_p)$ which
depends explicitly on the proper distance $r_p$ but not on the time
$t$.

Under conventional lensing conditions, a lightray spends a very small
amount of time in the area of influence of the lens (where
$\varphi\neq 0$). During such a small time the scale factor of the
universe $a(t)$ changes very little. It is typically assumed that
under such conditions, this scenario is consistent with the Einstein
equations for a linearized perturbation off a FRW spacetime.  The
consistency of a general cosmological lensing scenario with the
Einstein equations is studied in \citet{futamase}.

By Eq.~\ref{poisson}, the potential $\varphi$ is an explicit
function of the proper distance $r_p$ but not the time $t$.
Necessarily, thus,  the potential is a function of both the comoving
distance $r$ and the coordinate time $t$ via:

\begin{equation} \varphi = \varphi(r_p) = \varphi(a(t) r) =
\varphi(t,r). \label{pot1} \end{equation}

\noindent Consequently, the metric, Eq.~\ref{m1}, is not strictly
static in the comoving coordinate system. This issue is somewhat
obscure in the standard references, including the excellent books by
\citet{ehlers} and \citet{petters}, where the potential is said to
be ``time independent.''  The time-variation of the potential is
small, however, in the conditions of lensing, where the scale factor
changes very little during the passage of a lightray (one has
$d\varphi/dt = (da/dt)\times(r/a)\times \partial\varphi/\partial
r$). So during the passage through the lens one may approximate the
scale factor by its value at some time $t_l$ representative of the
time at which the lightray passes the lens.  This would make the
metric around the lens approximately static for the purposes of
calculating, for instance, the time delay due to the lens.

For our purposes the metric is needed for the entire trajectory from
the source to the lens, along which one expects the scale factor to
change perhaps significantly. The metric is not, thus, independent
of time, but it is conformal to an approximately time-independent
metric. Since the null geodesics of conformally related spacetimes
are identical, we may choose to approximate the scale factor with
its value at the lens or not. Although the approximation is very
valuable for the purpose of obtaining a closed form expression for
the lens equation in the conventional approach, it brings no real
advantage to the numerical integration of the null geodesics, so we
prefer to maintain the time dependence as prescribed by Poisson's
equation. The equations of motion are more complicated (in
particular, they do not decouple) but the added complications do not
represent a real obstacle.

Since the lens model Eq.~\ref{m1} is spherically symmetric, the
particle trajectories are planar, and there is no loss of generality
in choosing the plane as $\theta = \pi/2$. The geodesic equations can
be found explicitly as the Euler-Lagrange equations of the Lagrangian

\begin{equation}  {\mathcal{L}} = (1+2 \varphi)
\dot t^2 - a^2(t) (1-2\varphi) \{ \dot r^2 + r^2 \dot \phi^2 \}
 = 0 , \label{lagrangian} \end{equation}

\noindent to first order in $\varphi$.  The Lagrangian is equal to
zero because the geodesics are null.

Since the coordinate $\phi$ is cyclic, the Euler-Lagrange equations
are equivalent to five first-order ODEs, which we can write as

\begin{eqnarray}  \dot t &=& v_t \nonumber \\
\dot r & = & v_r \nonumber\\
\dot v_t & = &
\left( 4 \varphi \frac{\partial \varphi}{\partial t} - \frac{da}{dt}
\frac{1}{a} \right) v_t^2 - 2(1-2\varphi) \frac{\partial
\varphi}{\partial r} v_t v_r \nonumber\\
\dot v_r &=& -2 v_t v_r \left(\frac{\frac{da}{dt}}
{a} - (1+2\varphi) \frac{\partial \varphi}{\partial t} \right)
\nonumber\\
&& + 4
\frac{\varphi}{a^2} \frac{\partial \varphi}{\partial r} v_t^2 -
\frac{2b^2}{a^4 r^2} \frac{\partial \varphi}{\partial r} (1+
6\varphi) \nonumber\\
&& + \frac{b^2}{a^4 r^3}(1+4\varphi) \nonumber \\\dot \phi
&=& -\frac{b}{a^2 r^2} (1+ 2\varphi). \label{ODES} \end{eqnarray}

\noindent  These equations are obtained by working to first order in
$\varphi$ and making use of ${\mathcal{L}} = 0$ in the form

\begin{equation} a^2 (\dot r^2 + r^2 \dot \phi^2) = (1+4\varphi)
\dot t^2. \label{r1} \end{equation}

The parameter $b$ arises from $\ddot \phi = 0$ and is related to the
``observation angle'' at the observer or the angle between the lightray
and the optical (radial) axis connecting the observer to the lens.  The
relationship is determined by taking the dot product of the spatial
part of the null vector at the observer with a unit
vector pointing towards the origin.  Using the spatial part of the
metric, Eq.~\ref{m1}, one obtains

\begin{equation} \sin\theta_{obs} = \theta_{obs} =
\frac{b}{r_0}(1+2\varphi_0), \label{ring} \end{equation}

\noindent where the potential, $\varphi_0$, is evaluated at the
observer position at $t_0$.

We will be integrating past null geodesics from the observer back to
hypothetical sources.  Therefore, we need to specify five constants
of integration:

\begin{equation} t_0, r_0, \phi_0, {v_t}_0, {v_r}_0. \label{constants}
\end{equation}

\noindent This is in addition to the constant $b$ which essentially
fixes the initial value of $\dot \phi_0$.  For the initial position
$(t_0, r_0, \phi_0)$ we take $t_0 = 1$ by setting the unit scale to
the age of the universe, $\phi_0 = 0$ by arbitrary choice, and set
$r_0$ using the redshift of the lens (which is placed at the
origin), as was described in section~\ref{back:sec}.  The initial
value of $\dot v_t$ is set to $-1$, and then by ${\mathcal{L}}=0$,
we determine the initial value of $v_r$.  Note that by choice of
units, we have $a(t_0) = 1$. From Eq.~\ref{r1} and \ref{ODES}, we
have

\begin{equation} {v_r}_0 = - \sqrt{(1+4\varphi_0) \times (1- b^2 /
r_0^2)}, \label{vr0} \end{equation}

\noindent where $\varphi_0 = \varphi(t=1,r=r_0)$, and we take the
overall minus sign to make the rays approach the lens.


\section{SIS and NFW Models} \label{models:sec}

The SIS and NFW matter distributions are simple
spherically-symmetric matter models in common use in lensing and
other studies. An inconvenience of both these models is that the
total mass is unbounded.  In this section, we develop the
three-dimensional gravitational potentials for these models assuming
that the matter extends to some proper radius $r_c$ with zero matter
density for $r_p>r_c$.  We also develop two-dimensional (projected)
thin-lens matter distributions for each model with and without
truncation for $r_p>r_c$.

\subsection{SIS Newtonian potential}

The singular isothermal sphere (SIS) model is appealing because it
predicts flat rotation curves.  In the
SIS model the mass enclosed in a sphere of proper radius $r_p$ is

\begin{equation} M(r_p) = \frac{2 \sigma_v^2 r_p}{G}, \label{sismass}
\end{equation}

\noindent where the parameter $\sigma_v$ is the velocity dispersion,
independent of the cluster redshift.  For $r_p<r_c$ where $r_c$ is some
truncation radius to be determined, the SIS proper mass density
$\rho_p^{SIS}$ is given by

\begin{equation} \rho_p^{SIS} = \frac{\sigma_v^2} {2 \pi G r_p^2}.
\label{sisrho} \end{equation}

We assume that the matter density of the perturbation vanishes for
$r_p>r_c$. Notice that the value of $r_c$ can be chosen independent
of redshift, because the SIS mass model contains no redshift
dependence. Thus for $r_p>r_c$ the potential is that of a point mass
enclosing a mass $M(r_c)$.

The potential for $r_p<r_c$ is the solution of Poisson's equation
that matches the point mass potential smoothly (continuously and
with continuous first radial derivative) at $r_c$. We thus have:

\begin{equation} \varphi(r_p) = \left\{ \begin{array}{ccr} 2
\sigma_v^2 \ln x - 2 \sigma_v^2 & \quad\quad & x<1 \\
-\frac{2\sigma_v^2}{x} & \quad\quad & x>1 \end{array} \right. ,
\label{potential_sis} \end{equation}

\noindent where $x = r_p/r_c$ is a natural dimensionless radial
parameter.

To have the potential in the comoving coordinates, we make the
substitution $r_p = a(t) r$ in the potential,
Eq.~\ref{potential_sis}. Henceforward, $x$ is to be thought of as $x
= \frac{a(t) r}{{r_c}}$. The three-dimensional SIS model is thus a
two-parameter model, depending on $\sigma_v$ and $r_c$.

\subsection{SIS Thin lenses}

In this subsection, we derive the projected mass density and a thin lens
equation for the SIS model with a truncation radius, and show how the
usual presentation arises as a limiting case in which one of the two
parameters is lost.

The projected two-dimensional mass distribution is

\begin{equation} \Sigma_p^{SIS} = \int_{-z_c}^{z_c} \,
\rho_p^{SIS}\, dz_p, \label{sig1} \end{equation}

\noindent with $\rho_p^{SIS}$ given by Eq.~\ref{sisrho} and where
$z_p$ is the proper coordinate along the optical axis. The limit of
integration is $z_c = \sqrt{r_c^2-s^2}$, where $s$ is the proper
distance to the optical axis. The integral can be evaluated in
closed form, yielding

\begin{equation} \Sigma_p^{SIS}(s) = \frac{\sigma_v^2}{\pi \,G \, s} \, \arctan\left(
\sqrt{\frac{r_c^2}{s^2} - 1} \right) .
\label{sig2} \end{equation}

\noindent The mass interior to a proper radius $s$ is
\begin{equation}
M(s) = \frac{2\sigma_v^2 s}{G} A(r_c/s)\label{mass1}
\end{equation}

\noindent with
\begin{equation} A(r_c/s) \equiv  \arctan\left(
\sqrt{\frac{r_c^2}{s^2} - 1} \right) + \frac{r_c}{s} \left( 1
- \sqrt{1 - \frac{s^2}{r_c^2}} \right) . \label{A}
\end{equation}

The bending angle as a function of $s$ is \citep{ehlers}

\begin{equation} \hat \alpha = \frac{4 G M(s)}{s}.
\label{bending_general} \end{equation}

\noindent For a source at position $y$ from the optical axis, the images
will be seen on the same line at positions $x$ related to $y$ by the lens
equation which, with Eq.~\ref{bending_general}, takes the form

\begin{equation} y = \frac{D_s}{D_l} x - 8 D_{ls} \sigma_v^2
\frac{x}{|x|} \, A(|x|/r_c) \label{tl1} \end{equation}

\noindent for angular diameter distances to the source, lens, and between
the lens and source, $D_s$, $D_l$, and $D_{ls}$, respectively. The
dependence on the truncation radius $r_c$ appears in the factor $A(|x|/r_c)$.
In typical lensing situations the impact parameter $|x|$ is much smaller than
the truncation radius $r_c$.  The Taylor expansion of  $A(|x|/r_c)$ for small
ratio $|x|/r_c$ is
\begin{equation}
A(|x|/r_c) = \frac{\pi}{2} - \frac{|x|}{2r_c} -
\frac{|x|^3}{24 r_c^3} + \ldots
\end{equation}

\noindent In practice, it is customary to neglect the dependence on the
truncation radius $r_c$ and substitute the factor $A(|x|/r_c)$ by its limiting
value $\pi/2$, leading to the standard SIS lens equation found for instance
in \cite{ehlers}. The comparison with conventional results is easier to make
by introducing scaled quantities,

\[ \bar{y} = \frac{y D_l}{\xi_0 \, D_s} \quad\&\quad \bar{x} =
\frac{x}{\xi_0}, \]

\noindent with

\begin{equation} \xi_0 = 4 \pi \left(\frac{\sigma_v}{c}\right)^2
\frac{D_l \, D_{ls}}{D_s}. \label{scalerad} \end{equation}

\noindent Using these scaled quantities, the lens equation reads

\begin{equation} \bar{y} = \bar{x} - \frac{2}{\pi}
\frac{\bar{x}}{|\bar{x}|} A(\bar{r}_c/|\bar{x}|),
\label{t3} \end{equation}

\noindent which in the limit of large truncation radius reduces to the
conventional form

\begin{equation} \bar{y} = \bar{x} - \frac{\bar{x}}{|\bar{x}|}. \label{tl4}
\end{equation}

One sees that the thin-lens SIS model really contains two
parameters.  Under the normal conditions of gravitational lensing
this is considered as a one-parameter model, because an assumption
has been made on the relative sizes of $r_c$ and $x$. In the
comparisons that follow, we will be making reference to both SIS
lens equations: the conventional one-parameter SIS model
(Eq.~\ref{tl4}), and the two-parameter SIS model (Eq.~\ref{t3}).

We will be interested in the Einstein ring radius, which is given by
the value of $x$ satisfying Eq.~\ref{tl1} with $y = 0$.  For the
conventional one-parameter thin-lens SIS model, the Einstein ring
angle $\theta_E^{tl} = x/D_l$ is given by

\begin{equation} \theta_E^{tl} = 4\pi\frac{\sigma_v^2}{c^2}
\frac{D_{ls}}{D_s}, \label{tl_ring} \end{equation}

\noindent Technically, this thin lens result assumes that the truncation
radius $r_c$ is large compared to the scale radius $\xi_0$ \citep{ehlers}.

In the two-parameter thin-lens SIS model model, the Einstein ring
angle is given by the root of

\begin{equation} \theta - 8\frac{D_{ls}}{D_s} \frac{\sigma_v^2}{c^2}
A = 0, \label{tl2_ring} \end{equation}

\noindent where $A$, by Eq.~\ref{A}, is

\begin{eqnarray} A &=& \arctan\left(
\sqrt{\frac{r_c^2}{D_l^2\theta^2} - 1} \right) \nonumber\\
&& + \frac{r_c}{D_l
\theta} \left( 1 - \sqrt{1 - \frac{D_l^2 \theta^2}{r_c^2}} \right).
\label{A2}
\end{eqnarray}

\subsection{NFW Newtonian potential}

The matter distribution of the NFW model \cite{NFW} is given by

\begin{equation} \rho_p = \frac{\delta_c\rho_cr_s^3}{r_p(r_p+r_s)^2},
\label{rho_nfw}
\end{equation}

\noindent where $\rho_c = 3 H^2(z)/(8\pi G)$ is the critical density
of the universe as a function of redshift.  The two parameters of
the model are the scale radius, $r_s$, and $\delta_c$. In the
literature $\delta_c$ is often replaced by the concentration
parameter $c$ defined by

\begin{equation} \delta_c = \frac{200}{3}
\frac{c^3}{\ln(1+c)-c/(1+c)} ,
\end{equation}

\noindent and the scale radius $r_s$ is replaced by the virial radius
$r_{200} = c \, r_s$, which is the radius inside which the average mass
density is $200 \rho_c$.

Like the SIS model, the total mass over all space is undefined, so we must
truncate the model at some radius $r_c$.  However, because the NFW model is
indexed to $\rho_c$, the mass interior to a constant proper radius varies
with redshift.

We choose to hold the mass of the halo constant and truncate the NFW
profile at $r_{200}$, so that the constant total mass of the halo is

\begin{equation} M_{200} = \frac{800\,\pi}{3} \rho_c r_{200}^3
 . \label{M200} \end{equation}

\noindent The truncation radius $r_c = r_{200}$ is then a known function
of redshift through $\rho_c$:

\begin{equation}  r_{200} = \left( \frac{3\,  M_{200}}{800 \pi \, \rho_c}
\right)^{1/3}. \label{r200c} \end{equation}

\noindent In practice, we pick a value for $r_{200}$ at the
time the lightray passes the lens and use Eq.~\ref{M200} to compute
constant halo mass and Eq.~\ref{r200c} to determine the truncation
radius as a function of time.  For $x \equiv r_p/r_{200} < 1$, the
enclosed mass is

\begin{equation} M = M_{200} \left( \frac{ \ln(1+cx) -
\frac{cx}{(1+cx)}} { \ln(1+c) - \frac{c}{(1+c)} } \right) ,
\label{Mnfw} \end{equation}

\noindent where $c$ is the concentration parameter.

The NFW gravitational potential is obtained by smoothly matching the
spherically symmetric gravitational potential for the mass density given by
Eq.~\ref{rho_nfw} to a point mass potential at $x=1$.  The result is

\begin{equation} \varphi = -\frac{G \, M_{200}}{r_{200}} \left\{ \begin{array} {ll}
 \frac{ \displaystyle{\frac{1}{x}}\ln (1+cx) - \displaystyle{\frac{c}{1+c} }} {
\displaystyle{\ln(1+c)} - \displaystyle{\frac{c}{1+c}}}  & x<1\\
&\\
\displaystyle{\frac{1}{x}} & x>1
\end{array}\right.
\label{nfw_pot1} \end{equation}

\noindent where for cosmological spacetimes, we take $x = a(t)
r/r_{200}$, with $r_{200}$ defined in Eq.~\ref{r200c} and constant
concentration parameter $c$.

\subsection{NFW Thin lenses} \label{nfw_thin:sec}

We begin our discussion of NFW thin-lens models by determining the
two-dimensional, projected mass density used in gravitational
lensing with an arbitrary truncation radius $r_c$.  We
initially scale by $r_s$ so that $\tilde r_p = r_p/r_s$ and use
cylindrical coordinates where $\tilde r_p = \sqrt{\tilde x^2+ \tilde
z^2}$ where $\tilde z = z/r_s$ runs along the optical axis and
$\tilde x = s/r_s$ is the scaled distance from the $z$ axis.  We
define $\gamma = r_c/r_s$ to be an arbitrary scaled cutoff radius
(we will later set $r_c = r_{200}$).

The surface mass density is then given by

\begin{eqnarray}
\Sigma_p^{NFW} \!\!&\equiv & \!\!2r_s\int_0^{\sqrt{\gamma^2- \tilde x^2}}
\!\!\!\!\rho d\tilde z \nonumber\\
&=& \!\!\!\!2\delta_c\rho_c r_s\int_0^{\sqrt{\gamma^2-\tilde x^2}}
\!\!\!\!\!\!\!\!\!\!\!\!\!
    \frac{d \tilde z}{\sqrt{\tilde x^2+ \tilde z^2}(1+ \sqrt{\tilde x^2+\tilde z^2})^2}
    , \nonumber\\ \label{sigma_nfw1}
\end{eqnarray}

\noindent and we make the change of variable $\tilde z=\tilde x\tan
u$. Then the closed form expressions for the projected mass density
are for $\tilde x>1$,

\begin{eqnarray}   \Sigma_p^{NFW} &= &2\delta_c\rho_cr_s
    \frac{1}{\tilde x^2-1}\Bigg[
    \frac{2\tilde x Y}{(1+\tilde x+Y^2(\tilde x-1))}
\nonumber\\ &&
    -\frac{2}{\sqrt{\tilde x^2-1}}
\arctan\left(Y\sqrt{\frac{\tilde x-1}{\tilde x+1}}\right)\Bigg],
\nonumber\\
\end{eqnarray}

\noindent and for $\tilde x<1$,

\begin{eqnarray}   \Sigma_p^{NFW} &=&
2\delta_c\rho_cr_s
    \frac{1}{\tilde x^2-1}\Bigg[
    \frac{2\tilde x Y}{(1+\tilde x+Y^2(\tilde x-1))} \nonumber\\ &&
    -\frac{2}{\sqrt{1-\tilde x^2}}
\mbox{arctanh}\left(Y\sqrt{\frac{1-\tilde x}{\tilde
x+1}}\right)\Bigg], \nonumber\\\label{sigma_nfw}
\end{eqnarray}

\noindent and for $\tilde x=1$,

\begin{equation}   \Sigma_p^{NFW} = 2\delta_c\rho_cr_s
    \frac{Y}{2}\left(1-\frac{Y^2}{3}\right),
\end{equation}

\noindent where

\begin{equation} Y \equiv \tan \left[
\frac12\arctan\sqrt{(\gamma/ \tilde x)^2-1}\right]. \end{equation}

\noindent As $\gamma = r_c/r_s \to \infty$ we have $Y\to 1$ and we
recover the conventional NFW projected matter density, independent
of the truncation radius, Eq (11) in \citet{wright}.

In the limit that $\gamma = r_c/r_s \to \infty$, a closed form
expression for the mass interior to a radius $\rho$ can be found and
is reported in \citet{wright} and elsewhere.  Written in terms of
the concentration parameter $c$ and the scaled projected radius in
the lens plane $\bar{s} = s/r_{200}$, this closed form expression for
the mass is

\begin{equation}  M(s) = \frac{100\, r_{200}^3 \,H^2}{\ln(1+c)-c/(1+c)}
\,  Q(\bar{s}), \label{thin_m_nfw1} \end{equation}

\noindent where $Q(\bar{s})$ is given by

\begin{equation} Q(\bar{s})
= \left\{ \begin{array}{ccr} \frac{2}{\sqrt{1-(c\bar{s})^2}}
\mbox{arctanh}{\sqrt{\frac{1-c\bar{s}}{1+c\bar{s}}}}
+ \ln{\frac{c\bar{s}}{2}} & \quad\quad & c \bar{s}<1 \\
 \frac{2}{\sqrt{(c\bar{s})^2-1}}
\arctan{\sqrt{\frac{c\bar{s}-1}{c\bar{s}+1}}}
+ \ln{\frac{c\bar{s}}{2}} & \quad\quad &
c \bar{s}>1 \end{array} \right. . \label{Q_term}
\end{equation}

There is no closed form expression that we are aware of for the mass interior
to a radius $s$ in the case of a truncated NFW model. We simply find the mass
by numerically integrating Eq.~\ref{sigma_nfw}. This numerical
integration can be carried out using Gauss-Legendre quadratures, since the
integrand is regular, or by turning the integral equation into a differential
equation and solving with a numerical ODE solver with adaptive stepsize. In
either case, it is critical to sample well the region near $s=0$ because
the integrand has derivatives that are undefined there.

Given the mass within a radius $|x|$, the lens equation
\begin{equation}
y = \frac{D_s}{D_l} x - \frac{4GD_{ls}M(|x|)}{x}\label{lens_general}
\end{equation}

\noindent is the basis for conventional gravitational lensing by NFW
models, in the same way as for the SIS model.  The Einstein ring
angles will be determined by the roots of Eq.~\ref{lens_general} for
$y= 0$.


\section{Experimental testbed} \label{sim:sec}

We maintain the source, lens and observer on the same axis, so that
the observer will see Einstein rings around the lens at an angle
$\theta_{obs}$ in Eq.~\ref{ring}. This observation angle will be
considered as a function of the lens model parameters and the
relative positioning of the lens, observer and source.

Since the SIS model is a one parameter model (holding fixed the truncation
radius $r_c$), fixing the Einstein ring angle results in a determined value
of $\sigma_v$.  It is fairly common in strong lensing articles to use a
simple SIS model in this manner, even when two symmetric arcs are found
instead of a ring.  While it is known that this method leads to an
overestimate of the lens mass, this direct relation between $\sigma_v$ and
the ring angle serves as a good comparison for this paper.

There is a degeneracy in specifying the Einstein ring angle as the only
observable for NFW models, since the NFW models contain two parameters.
However, weak lensing studies based on gravitational shearing of background
images tend to show that the virial radius $r_{200}$ is often better
constrained than the concentration parameter $c$.  In addition, weak lensing
studies tend to find very little tangential shear, a direct measure of the
mass, at physical radii greater $3.5$~Mpc.  For these reasons, in this paper
we will generally specify that $r_{200} = 3.5$~Mpc when the light passes the
lens and think of the concentration parameter of the NFW models as the
dynamical variable tied to observed Einstein rings. We find that the errors
in the thin-lens cited in this paper do not appreciably vary when $r_{200}$
is reset to new values within an expected $r_{200}$ range.

We consider a number of galaxy clusters with known redshifts as lens
candidates.  As a benchmark system, we consider RXJ1347-1145, a high
X-ray luminosity cluster at $z=0.45$ that has been widely studied in
weak and strong lensing.  RXJ1347-1145 has a pair of arcs at
redshift 0.8, located at approximated 35 arc sec from the
gravitational center of the cluster.  We also consider a cluster
discovered by \citet{wittman} at $z=0.68$ that appears to have a
pair of arcs at 7 arc sec at an unconfirmed redshift.  To consider a
range of lens redshifts, we also include Abel 1451 at $z=0.2$, whose
weak lensing signal is reported in \citet{cypriano} and
RDCS1252.9-2927, a high redshift cluster ($z=1.24$) with a weak
lensing measurement from HST imaging reported in \citet{lombardi}.

We note that the mass density inferred from the weak lensing measurements of
these clusters has fallen to essentially zero at a projected radius in the
lens plane by approximately 3.5 Mpc. Therefore, in our standard comparisons,
we choose the SIS truncation radii of $3.5$~Mpc, the same as the NFW cases,
although in some comparisons we allow for a range of truncation radii.
$3.5$~Mpc is generally a factor of 10 to 50 larger than the scale radius,
Eq.~\ref{scalerad}.


\section{Numerical approach and accuracy}

For the purposes of integrating the null geodesics, we re-scale the time and
radial coordinates:

\begin{equation} t' = \frac{t}{t_o} \quad\quad r' = \frac{r}{ct_o}.
\label{c1} \end{equation}

\noindent In the $t'$ coordinate, the observer receives the light rays
$t' = 1$, and the metric is rescaled by an overall factor

\begin{eqnarray} ds^2 &=& (ct_o)^2\Big(dt'^2 - a^2(t')\nonumber\\
&&\times \{ dr'^2 + r'^2 (d\theta^2
+ \sin^2\theta d\phi^2) \}\Big), \hspace{0.5cm}\label{FRW2} \end{eqnarray}

\noindent which, of course, does not alter the null geodesics.

We use a ray shooting technique to determine the observation angle
for a given positioning and set of lens model parameters.  For
example, if the Einstein ring angle is fixed, increasing $\sigma_v$
or $c$ will cause the true light ray to cross the optical axis
closer to the lens.  This allows us to vary one model parameter at a
time for a given Einstein ring angle, using Newton's method to
determine the parameter value to a high accuracy, limited
essentially by the quality of the ODE integration scheme.

We integrate the null geodesic equations using an adaptive stepsize
Runge-Kutta-Fehlberg 4-5 method based on the implementation in
\citet{nrc}.  This allows us to monitor the error in each step and
maintain a known, and small, accumulated error in the integration.
We purposefully slow the integration (over the affine parameter) as
we approach the source location, where the adaptive stepsize
algorithm would naturally take large steps, in order to carefully
stop at the source position.

The principal source of error in our method is in stopping at the
source position.  With the observer at $\phi=0$ and source at
$\phi_f$, we integrate over the affine parameter until $\phi =
\phi_f \pm \epsilon_\phi$ and determine whether $r = r_l \pm
\epsilon_r$ where $r_l$ is the known lens position.  Based on
whether $r$ is less than or greater than $r_l$, we accept or change
initial values at the observer.  In practice, we find that the error
introduced by this stopping condition is orders of magnitude greater
than the error accumulated in the ODE integration.  In the materials
below, the error bars on the numerical integration are derived from
this known error source.

We consider results from the numerical integration of the null
geodesic equations, Eq.~\ref{ODES}, to be the ``correct'' results
from lensing which we compare the thin-lens models to.  As an
estimate of the error in our ODE integration, we first set the
Einstein ring angle to known value, $\theta_E$, and solve for the
parameter value (either $\sigma_v$ or $c$).  We then solve backwards
for the Einstein ring angle our numerical integration predicts given
these parameter values, and subtract from the original specified
angle.

\begin{figure}
\plotone{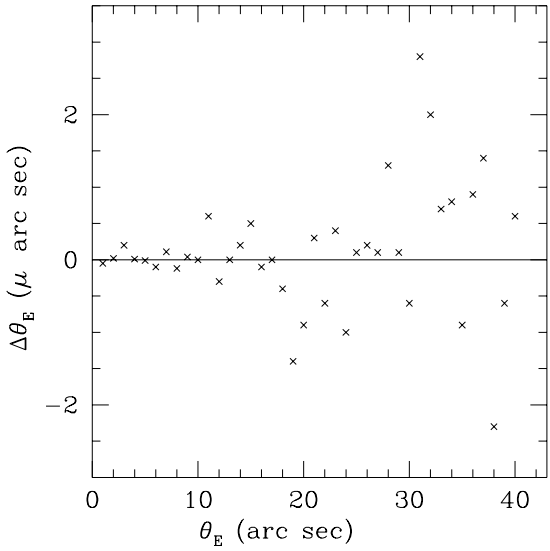} \caption{\label{consist_n_sis:fig} A plot
estimating the numerical error introduced by integrating the null
geodesics and using ray shooting to determine the velocity
dispersion (SIS model) or Einstein ring angle.  We assume a
three-dimensional SIS lens with matter extending to a co-moving
radius of 3.5 Mpc lying at $z=0.45$ and a source at $z=0.8$. The
Einstein ring angle is set, then the velocity dispersion is
determined.  From this velocity dispersion, we recalculate the
Einstein ring angle and subtract from the original angle.  These
differences are all close to zero and show no trend.}
\end{figure}

The difference, which should be zero, is an estimate of the error
our methods allow.  Figure~\ref{consist_n_sis:fig} shows this error
estimate for a SIS model of RXJ1347-1145, a cluster at $z=0.45$ with
arcs from a $z=0.8$ source. The error values are all small and show
relatively little trend. Similar results are obtained in the NFW
models.

By contrast, when one specifies the Einstein ring angle, solves the
geodesic equations to determine the model parameters, then asks for
the value of the ring angle predicted by the thin-lens model for
those parameters, one sees a significant difference from the
original angle.  Figure~\ref{consist45:fig} shows the error, $\Delta
\theta_E = \theta_E - \theta_E^{tl}$, in the Einstein ring angle
predicted by the thin-lens methods for a SIS model to be much larger
than the numerical scatter for RXJ1347-1145.

\begin{figure}
\plotone{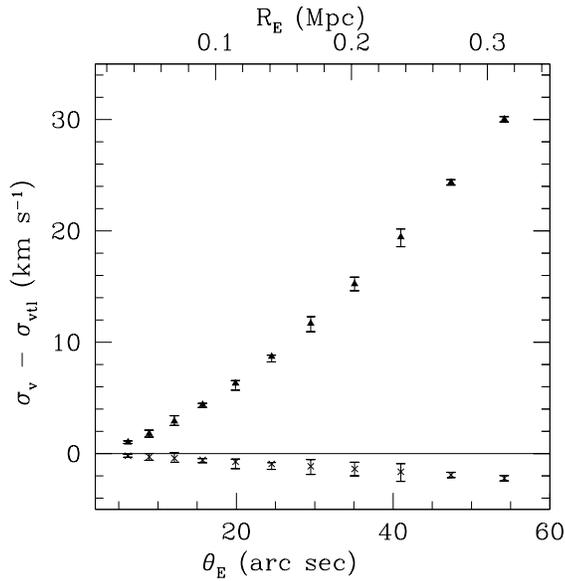} \caption{\label{sis_err:fig} A plot of the error in
the velocity dispersion introduced by two thin-lens models for arcs
of a three-dimensional SIS whose matter extends to a co-moving
radius of 3.5 Mpc. The error bars here are drawn 1000 times larger
than the actual error bars and reflects an overestimate of the
accumulated error in the numerical integration. The triangles
correspond to the usual thin-lens SIS model, and the crosses
correspond to a thin-lens SIS model that accounts for the finite
radius.}
\end{figure}

Using the error introduced by the stopping conditions to form error
bars, we show the difference between the predicted velocity
dispersion for a given angle in a SIS model in
Fig.~\ref{sis_err:fig} for a RXJ1347-1145 setup with $r_c =
3.5$~Mpc. The error bars are drawn $1000$ times larger than the
actual error.

These error estimates and plots with error bars confirm that the
differences measured between values predicted by the numerical
integration of the geodesic equations and the values predicted by
the various thin-lens models are not attributable to truncation or
roundoff error in the numerical code.


\section{Thin-lens accuracy} \label{compare:sec}

Since we are primarily interested in the error introduced in the
thin-lens approximation in predicting the mass of the cluster, we
cite the relative error in the square of the velocity dispersion,

\begin{equation} \frac{\Delta \sigma_v^2}{\sigma_v^2}
= \frac{\sigma_v^2 - {\sigma_v^{tl}}^2} {\sigma_v^2},
\label{sis_err}
\end{equation}

\noindent for SIS models, and the relative error in $c$,

\begin{equation} \frac{\Delta c}{c}
= \frac{c - c^{tl}} {c}, \label{nfw_err}
\end{equation}

\noindent for NFW models.  The SIS error is equal to the fractional
error in the predicted mass.  However, due to the differences in the
NFW 3-d enclosed mass and conventional enclosed masses, it is more
difficult to think of errors in $c$ translated to errors in mass for
NFW models.

As a first comparison, we consider a lens and source for
RXJ1347-1145 (source at $z=0.8$) and RDCS1252.9-2927, where we
assume a source is located at $z=1.5$ on the optical axis, and
compute the velocity dispersion for a given observed Einstein ring
angle using the non-perturbative method and the two thin lens
methods. Figure \ref{vtheta35:fig} shows the relative error in the
square of the velocity dispersion as a function of the observed
Einstein ring angle for a cutoff radius of 3.5 Mpc. We see that the
errors tend to approach $1\%$ for physical values for the
conventional one-parameter SIS thin-lens.  The truncated SIS thin
lens models tend to perform better, with approximately one fourth
the fractional error in the total cluster mass.

\begin{figure}
\plotone{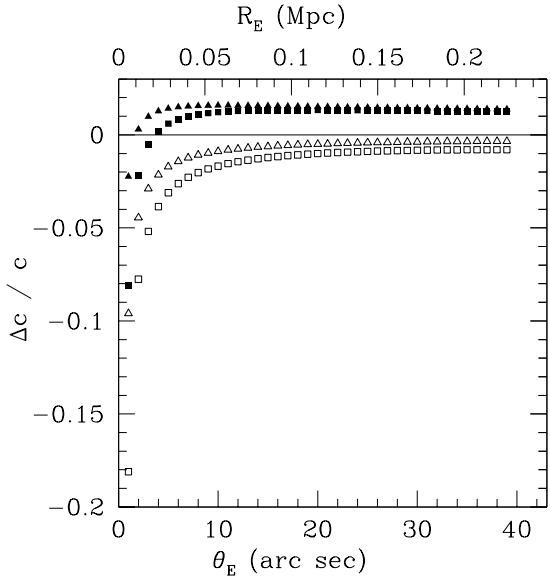} \caption{\label{dcerr2:fig} A plot of the relative
error in $c$ for an NFW model as a function of the observed Einstein
ring angle.  The triangles and squares correspond to lens / source
redshifts of (0.45, 0.8) and (1.24, 1.5) respectively.  The open
symbols are the relative error in the NFW model with a cutoff
radius, and the closed symbols show the error with no cutoff radius.
The cutoff radius used is set to the virial radius ($r_{200}$) as
3.5 Mpc at the lens crossing time and then varies to maintain a
constant halo mass. The Mpc scale corresponds to the projected
Einstein ring radius for a lens at $0.45$; the projected radius for
a lens at $1.24$ would be larger.}
\end{figure}

For the same lens and source combinations, Fig.~\ref{dcerr2:fig}
shows the relative error in $c$ for an NFW model given the Einstein
ring angle.  Here we assume that $r_{200} = 3.5$~Mpc when the light
ray passes the lens, which is the best fit scale radius for
RXJ1347-1145 for weak lensing analysis from \citet{kling_rxj}.  We
see that the error is on the same order of magnitude as the error in
the SIS mass prediction for arcs at large observation angles. For
small observation angles, there is a significant difference between
the values for $c$ predicted by the non-perturbative and thin lens
models, but these observation angles are unlikely, because they
involve low concentration parameters ($c<1.0$). The truncated
thin-lens NFW models again perform better for physical systems than
the conventional thin-lens models, but the error remains on the same
order of magnitude.

\begin{figure}
\plotone{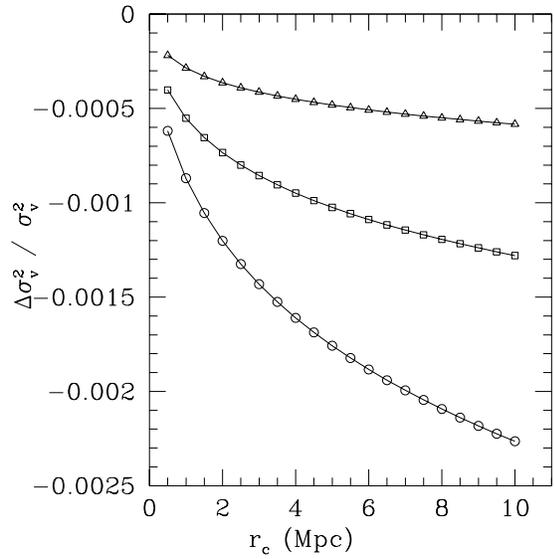} \caption{\label{vrcwith:fig} A plot of the relative
error in the square of the velocity dispersion for thin lens, SIS
with a cutoff radius as a function of the cutoff radius. The three
curves correspond to errors if the Einstein ring angle is 5
(triangles), 15 (squares) or 35 (circles) arc sec.  Here we are
considering lensing by RXJ1347-1145, which does have arcs at 35 arc
sec.}
\end{figure}

In Fig.~\ref{vrcwith:fig}, we show how the relative error in the
square of the velocity dispersion depends on the cutoff radius for
conventional and truncated SIS models, respectively.  Here, we
consider RXJ1347-1145, and the three curves represent Einstein rings
observed at 5, 15, and 35 arc sec.  We see that the relative error
introduced by using a thin-lens model does depend on the choice of
truncation radius and does not go to zero with larger radius.

For NFW models, variation in $r_{200}$ at the time the light ray
passes the lens leads to minor changes in the value of $c$, but not
significant changes in the error estimates.  For example, weak
lensing measurements found the best fit parameters for RXJ1347-1145
to be $r_{200} = 3.5_{-0.2}^{+0.8}$~Mpc \citep{kling_rxj}. For the
physical arcs in this system at 35 arc sec,
Table~\ref{nfw_r200:table} indicates that the fractional errors
maintain the same order of magnitude as $r_{200}$ is varied.

Table \ref{sis_35:table} gives the relative error in mass (or square
velocity dispersion) for thin-lens SIS models for four possible
scenarios at increasing lens redshift,  all with truncation radii of
3.5 Mpc.  In all four cases the lens and the source are at fixed
resfhit, and the Einstein ring angle is varied. A number of
observations can be made.  First, the error in the prediction of the
mass incurred by the use of the thin lens approximation alone varies
between $0.03\%$ for small rings and $0.3\%$ for large rings.
Secondly, the error in the mass incurred by the combination of the
thin lens approximation and the removal of truncation varies between
$0.1\%$ for small rings and $2\%$ for large rings.  Subtracting both
errors yields an estimate of the error introduced by the removal of
truncation in the thin lens model. One can see that the removal of
truncation along the line of sight introduces
an error between 5 and 12 times as large as the thin lens assumption
does, for small and large rings respectively.

Similarly, Table \ref{nfw_35:table} gives the relative error in $c$
for NFW models where the truncation radius is set to $r_{200} =
3.5$~Mpc when the light ray passes the lens.  The scenarios are the
same as for Table \ref{sis_35:table}.  One can see that the thin lens
approximation alone introduces and error in $c$ between $0.3\%$ and
$2\%$ in cases of lens redshift less than 1.  Perhaps unexpectedly,
the conventional approach shows a relative error around $1\%$ to
$2\%$ due to the combination of the thin lens assumption and removal
of truncation.  The truncation alone introduces an error between
between 1 and 5 times that incurred by the thin lens assumption.


\section{Cosmology Dependence} \label{cosmo:sec}

For a given lens and source distance, and a specified observation
angle, the values of parameters in a matter distribution model will
depend on the assumed cosmological parameters.  In this section, we
briefly show that while the values of the parameters vary with
choice of cosmology, the overall fractional uncertainty in those
parameters does not show significant variation.

In particular, because the NFW model is indexed to the critical
density as a function of redshift, one might worry that variation in
cosmology will have a strong effect on the accuracy of NFW model
parameters.

As a simple test case, we present two tables that show the relative
errors associated with SIS and NFW model parameters for the physical
arcs appearing at 35~arc sec in RXJ1347-1145 as a function of
cosmology.  While we only consider flat cosmological models, we do
consider a wide range in the matter density $\Omega_m$.

Tables \ref{cosmo_sis:table} and \ref{cosmo_nfw:table} present the
relative errors in $\sigma_v^2$ and $c$ respectively for a cutoff
radius of 3.5~Mpc.  For the NFW model, $r_{200}$ is set to the
cutoff radius as before.  Both tables show that the accuracy of thin
lens models changes only by a factor of 3 or so over a very wide
range in flat cosmologies.


\section{Discussion} \label{discussion:sec}

\begin{figure}
\plotone{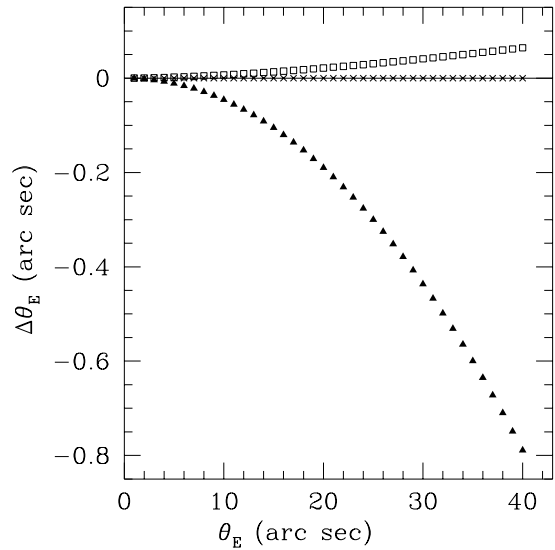} \caption{\label{consist45:fig} A plot showing the
numerical scatter of Fig.~\ref{consist_n_sis:fig} (crosses) and
differences between actual and thin-lens predicted Einstein ring
angles for SIS models with (boxes) and without (triangles) a cutoff
radius for a lens and source at z = 0.45 and 0.8 respectively.  A
cutoff radius of 3.5 Mpc is used. }
\end{figure}

Figure \ref{consist45:fig} shows comparisons of predicted Einstein
ring angles by conventional approaches and by properly truncated
thin-lenses, compared to the relativistic prediction by numerical
integration. One can see that given a fixed lens mass and distances
to the lens and source, just the use of a thin-lens approximation
results in a smaller Einstein ring angle for the same truncated mass
model.  From this, as well as from Tables  \ref{sis_35:table} and
\ref{nfw_35:table}, we conclude that the use of the thin-lens
assumption alone leads to an overestimate of the mass, at least in
the case of the two spherically symmetric models that we use.

In calculating the total bending angle accumulated along a photon
path, the thin-lens assumption leads directly to the substitution of
an undeflected path for the actual photon path, which, being curved,
is normally longer than the undeflected path. In all models where
the bending increases monotonically along the path, this procedure
naturally leads to an underestimate of the total bending, as the
integration takes place over a shorter interval.  To compensate for
the underestimate of bending, the thin-lens assumption overestimates
the mass.

In the conventional SIS model, setting the value of $\sigma_v$ the
same as the value in the truncated models essentially means that the
conventional model contains more mass: see Eq.~\ref{mass1}.  With
more mass, the Einstein ring angle grows, so the conventional
Einstein ring angles in Fig.~\ref{consist45:fig} are larger than the
correct angles.

\begin{figure}
\plotone{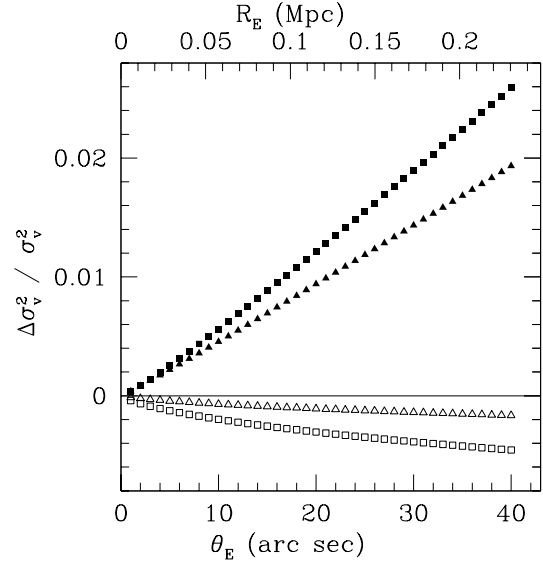} \caption{\label{vtheta35:fig} A plot of the
relative error in the square of the velocity dispersion for an SIS
model as a function of the observed Einstein ring angle.  The
triangles and squares correspond to lens / source redshifts of
(0.45, 0.8) and (1.24, 1.5) respectively.  The open symbols are the
relative error in the SIS model with a cutoff radius, and the closed
symbols show the error with no cutoff radius.  The cutoff radius
used is 3.5 Mpc. The Mpc scale corresponds to the projected Einstein
ring radius for a lens at $0.45$; the projected radius for a lens at
$1.24$ would be larger.}
\end{figure}

Figures \ref{sis_err:fig} and \ref{vtheta35:fig} indicate that for
the same observation angles and source and lens positions, the
inferred velocity dispersion for a truncated thin-lens model
overestimates the true velocity dispersion.  Since at the same mass,
the truncated thin-lens predicts a smaller angle, increasing the
angle to align with the true observation angle increases the mass of
the lens, and hence the velocity dispersion.

The conventional SIS can achieve this extra mass either by increasing
the velocity dispersion or by adding mass along the line of sight
past the truncation radius.  This second method can, in fact, drive
down the velocity dispersion.  In Fig.~\ref{vtheta35:fig}, we see a
lower velocity dispersion for the conventional model because the
extra mass has been stored along the line of sight.  For other
lens-source-ring angle configurations, plots similar to
Fig.~\ref{vtheta35:fig} show negative dips in the non-truncated
square velocity dispersion error at small observation angles, which
implies that the conventional SIS model has a higher than correct
velocity dispersion.

The relative error in the concentration parameter of NFW models,
Fig.~\ref{dcerr2:fig}, shows a similar effect.  Increasing the
concentration parameter makes the core of the NFW profile steeper,
placing more mass towards the center of the distribution. For the
truncated NFW thin lens, increasing the concentration parameter
increases the central mass to compensate for a required increased
bending angle in the same way it did in the SIS case.  For the
conventional NFW model, the trade-off between extra stored mass
along the line of sight and increasing the steepness of the profile
results in a shift in the sign of the relative error in $c$.


\section{Conclusion} \label{conclusion:sec}

The principal finding of this paper is that, in the context of the
current generation of precision cosmology experiments, the use of the
conventional approach in strong gravitational lensing is justified.
Our figures and tables demonstrate that in a wide variety of strong
lensing scenarios, and for two widely used mass models, the
conventional approach works well in predicting cluster masses. Except
in cases where the truncation radius is on the order of 1 to 2 Mpc,
the errors in the SIS predicted mass are generally less than $2\%$.
The error in concentration parameter for the NFW model is also small
except for very high redshift clusters.  This shows that concerns
over the possibility of artificially high estimates of dark matter
content in clusters as a result of the use of the thin-lens
approximation are generally unfounded.

For the first time, to our knowlegde, we have been able to show that
the thin lens assumption and the removal of the truncation radius
along the line of sight contribute to the error in very different
measures.  Interestingly, the two approximations introduce errors in
opposite directions: the thin-lens assumption underestimates the
mass, whereas the removal of the truncation radius overestimates it.
The error introduced by removing the truncation along the line of
sight is of the same order of magnitude as that of the thin-lens
assumption in the best case, and up to an order of magnitude larger
in the worst case within our study.   The truncation error is thus
the dominant source of  error in the conventional approach.  This is
a strong indication that, as precision increases in strong lensing
observations, in order to gain a significant increase in accuracy it
will be necessary, and perhaps even sufficient at first, simply to
recover the truncation radius as an additional parameter in thin-lens
models.

Truncation issues aside, we find that the error introduced by the
thin lens assumpion alone for a given truncation method (in our case,
hard truncation) varies between $0.03\%$ and $0.3\%$ in the SIS case,
and between $0.3\%$ and $2\%$ in the NFW case for lenses at redshift
less than 1.  Knowledge of the size of this error should be taken
into account when considering additional approximations such as
proper truncation methods.  Errors of this size may become
significant in the next 20 years as the observational precision
increases.

Perhaps more important than the specific strong lensing finding in
this paper is the possible implication of this finding for weak
lensing surveys of galaxy clusters.  Strong lensing depends of the
first derivative of the gravitational potential (the Christoffel
symbols of the underlying metric), but weak lensing depends on the
second derivative (the curvature).  One may advance the conjecture
that the higher order (derivative) measurements of weak lensing might
be less sensitive to the thin-lens approximation, and therefore more
accurate in the determination of the cluster mass.

At the time at which the fundamental parameters are known to $1\%$
or better, this paper develops the appropriate steps for pursuing
strong gravitational lensing without lens planes.  In particular, we
have demonstrated gravitational lensing by weak gravitational
potentials in the appropriate cosmological backgrounds by numerical
integration of the null geodesic equations of general relativity to
be a feasible task.

A long-standing argument for the use of the thin-lens approximation and
against integrating the null geodesic equations has always been
the high computational cost of numerical integration compared with
the low cost of the (essentially algebraic) thin-lens approach. Ten
years ago this was essentially true. However, the current speed and
memory capacity of even small workstations removes this argument
against using the general relativistic equations instead of the
thin-lens method or other higher order approximations such as those
proposed by \citet{pyne} or \citet{knp1}.

\acknowledgements

This material is based upon work supported by the National Science
Foundation under Grant Nos. PHY-0244752 and PHY-0555218, and a
Bridgewater State College Faculty and Librarian Research Grant. We
gratefully acknowledge the hospitality and support of the American
Institute of Mathematics during the progress of the workshop
``Gravitational Lensing in the Kerr Geometry,'' AIM, Palo Alto, July
5-10, 2005.

%
%

\clearpage

\begin{table}[hp] \begin{center}\begin{tabular}{ccc}
\tableline $r_{200}$ (Mpc) & $c$ & $\delta$  \\
\tableline ~&~&~\\
3.3 & 4.9 &  -0.003 \\
3.5 & 4.3 & -0.004 \\
4.3 & 2.8 & -0.006 \\
 \tableline
\end{tabular} \caption{Relative error in $c$ in NFW thin-lens models
with a truncation radius for three different values of $r_{200}$ as
the light ray passes the lens for the 35 arc sec arcs of.
\label{nfw_r200:table} }
\end{center}
\end{table}


\clearpage

\begin{table}[hp] \begin{center}\begin{tabular}{lrrrrcc}
\tableline System &  $\theta_E$ & $\sigma_v$ &
$\delta_1$ & $\delta_2$ & $
\delta_3$&$\delta_3/|\delta_2|$\\
\tableline ~&~&~&~&~&~&~\\

~ &  5 & 495.5 & 0.0012 & -0.00028& 0.0015 & 5 \\
~&  10 & 701.1 & 0.0026 & -0.00045 & 0.0031 & 7 \\
Abel 1451  & 15 & 859.3 & 0.0039 & -0.00059 & 0.0044 & 8 \\
$z_l=0.2$ & 25 & 1111 & 0.0067 & -0.00083 & 0.0075 & 9 \\
$z_s=0.8$  & 35 & 1316 & 0.0095 & -0.001 & 0.012 & 11 \\
\tableline
~ & 5 & 674.3 & 0.0022 & -0.00043  & 0.0026 & 6\\
~ & 10 & 954.7 & 0.0045 & -0.00069  & 0.0052 & 8\\
RXJ1347-1145&  15 & 1171 & 0.007 & -0.0009 & 0.0079 & 9 \\
$z_l=0.45$ &25 & 1515 & 0.012 & -0.0012  & 0.0013 & 11\\
$z_s=0.8$ &35 & 1797 & 0.017 & -0.0015  & 0.019 & 12\\
\tableline
~ &  5 & 818.7 & 0.0026 & -0.0006 & 0.0032 & 5 \\
~  &  10 & 1160 & 0.0055 & -0.00095 & 0.0064 & 7 \\
(Wittman) &  15 & 1422 & 0.0084 & -0.0012 & 0.0096 & 8 \\
$z_l=0.68$ &  25 & 1842 & 0.014 & -0.0017  & 0.016 & 9\\
$z_s=1$ &  35 & 2186 & 0.02 & -0.0021 & 0.022 & 11 \\
\tableline
~ &  2 & 767.1 & 0.00086 & -0.00066 & 0.0015 & 2 \\
~ &  4 & 1086 & 0.002 & -0.0011 & 0.0031 & 3 \\
RDCS 1252.9-292& 7 & 1437 & 0.0037 & -0.0016 & 0.0053 &
3 \\
$z_l=1.24$ & 10 & 1719 & 0.0056 & -0.002 & 0.0076 & 4 \\
$z_s=1.5$ &  15 & 2109 & 0.0088 & -0.0026 & 0.011 & 4 \\
\tableline
\end{tabular} \caption{Relative error in the predicted square
velocity dispersion for an SIS model with hard truncation at 3.5~Mpc. The hypothetical Einstein ring angles are given in arc sec,
and the velocity dispersions are given in km~s$^{-1}$. RXJ1347-1145
has arcs at approximately 35~arc sec, while Wittman has arcs at
7~arc sec.  The column headed by $\delta_1$ shows the error in the prediction of
$c$ by the model with no truncation radius along the line of sight.
Column $\delta_2$ is the error by the model with hard truncation, and
represents the error intruced by the thin-lens approximation alone.
The difference $\delta_3 \equiv\delta_1-\delta_2$ represents the error introduced by
the removal of the truncation radius.  The last column to the right
indicates the size of the truncation error compared to the size of the
thin-lens error.  \label{sis_35:table} }
\end{center}
\end{table}


\clearpage

\begin{table}[hp] \begin{center}\begin{tabular}{lrrrrrc}
\tableline System & $\theta_E$ & $c$ &
$\delta_1$ & $\delta_2$& $
\delta_3$&$\delta_3/|\delta_2|$ \\
\tableline ~&~&~&~&~&~&~\\

~ &  5 & 1.992 & 0.0043 & -0.019 & 0.023& 1\\
~ &  10 & 2.447 & 0.0084 & -0.0096 & 0.018& 2\\
Abel 1451  & 20 & 3.136 & 0.0095 & -0.0052 & 0.015& 3 \\
$z_l=0.2$ & 30 & 3.759 & 0.0096 & -0.0038& 0.013 & 4 \\
$z_s=0.8$ &  40 & 4.392 & 0.0094 & -0.0032 & 0.013 & 4 \\
\tableline
~ &  5 & 1.744 & 0.014 & -0.017& 0.032 & 2 \\
~ &  10 & 2.234 & 0.016 & -0.0088 & 0.025 & 3 \\
RXJ1347-1145 &  20 & 3.048 & 0.015 & -0.005  & 0.02& 4\\
$z_l=0.45$ &  30 & 3.879 & 0.014 & -0.0038  & 0.018 & 5\\
$z_s=0.8$&40 & 4.843 & 0.014 & -0.0033 & 0.017 & 5 \\
\tableline
~ &  5 & 1.64 & 0.017 & -0.019  & 0.036 & 2\\
~ &  10 & 2.145 & 0.018 & -0.01 & 0.028 & 3 \\
(Wittman) &  20 & 3.018 & 0.017 & -0.0058 & 0.023 & 4\\
$z_l=0.68$ &  30 & 3.964 & 0.016 & -0.0046 & 0.021& 4 \\
$z_s=1$ &  40 & 5.138 & 0.015 & -0.0042  & 0.019 & 5\\
\tableline
~ &  2 & 1.175 & -0.022 & -0.078  & 0.056 & 1\\
~ &  4 & 1.515 & 0.002 & -0.039 & 0.041 & 1\\
RDCS 1252.9-292 & 7 & 1.883 & 0.0099 & -0.023 & 0.033 & 1\\
$z_l=1.24$ & 10 & 2.198 & 0.012 & -0.01 & 0.029 & 2 \\
$z_s=1.5$ & 15 & 2.691 & 0.013 & -0.012 & 0.025 & 2 \\
\tableline
\end{tabular} \caption{Relative error in the concentration parameter,
$c$, for an NFW model. The truncation radius is set to the
co-moving physical distance value of $r_{200} = 3.5$~Mpc at the lens
crossing time and then varies to maintain a constant mass. The
hypothetical Einstein ring angles are given in arc sec. RXJ1347-1145
has arcs at approximately 35~arc sec, while Wittman has arcs at
7~arc sec. The column headed by $\delta_1$ shows the error in the prediction of
$c$ by the model with no truncation radius along the line of sight.
Column $\delta_2$ is the error by the model with hard truncation, and
represents the error intruced by the thin-lens approximation alone.
The difference $\delta_3 \equiv \delta_1-\delta_2$ represents the error introduced by
the removal of the truncation radius.  The last column to the right
indicates the size of the truncation error compared to the size of the
thin-lens error.
 \label{nfw_35:table} }
\end{center}
\end{table}


\clearpage

\begin{table}[hp] \begin{center}\begin{tabular}{ccccc}
\tableline $\Omega_m$ & $\Omega_\Lambda$ & $\sigma_v$ (km s$^{-1}$)
& $\delta_1$ & $\delta_2$ \\
\tableline ~&~&~&~&~\\

0.05 & 0.95 & 1708 & 0.019 & -0.0013 \\
0.15 & 0.85 & 1750 & 0.018 & -0.0014 \\
0.30 & 0.70 & 1797 & 0.017 & -0.0015 \\
0.45 & 0.55 & 1833 & 0.016 & -0.0016 \\
0.60 & 0.40 & 1862 & 0.015 & -0.0017 \\
0.95 & 0.05 & 1913 & 0.014 & -0.0018 \\

\tableline
\end{tabular} \caption{Relative error in the SIS predicted mass or
$\sigma_v^2$, for a SIS model thin lens without ($\delta_1$) and
with ($\delta_2$) a cutoff radius of 3.5~Mpc for the physical arcs
of RXJ1347-1145 (at 35 arc sec) for various flat cosmologies.
\label{cosmo_sis:table} }
\end{center}
\end{table}


\clearpage

\begin{table}[hp] \begin{center}\begin{tabular}{ccccc}
\tableline $\Omega_m$ & $\Omega_\Lambda$ & c
& $\delta_1$ & $\delta_2$ \\
\tableline ~&~&~&~&~\\

0.05 & 0.95 & 6.11 & 0.01 & -0.0023 \\
0.15 & 0.85 & 5.25 & 0.012 & -0.0028 \\
0.3 & 0.7 & 4.34 & 0.014 & -0.0035 \\
0.45 & 0.55 & 3.7 & 0.017 & -0.0044 \\
0.6 & 0.4 & 3.22 & 0.02 & -0.0053 \\
0.95 & 0.05 & 2.46 & 0.027 & -0.0079 \\

\tableline
\end{tabular} \caption{Relative error in the NFW predicted
concentration parameter, for an NFW model thin lens without
($\delta_1$) and with ($\delta_2$) a cutoff radius for various flat
cosmologies. The cutoff radius is set to the co-moving physical
distance value of $r_{200} = 3.5$~Mpc at the lens crossing time and
then varies to maintain a constant mass. \label{cosmo_nfw:table} }
\end{center}
\end{table}


\clearpage



\end{document}